\documentclass{aastex}
\usepackage{spr-astr-addons}
\usepackage{url}
\usepackage{graphicx}
\usepackage{longtable}
\usepackage{comment}
\usepackage{natbib}
\usepackage{verbatim}
\usepackage{morefloats}
\usepackage{lscape}
\usepackage{threeparttable}
\usepackage{wasysym}
\usepackage{txfonts}
\usepackage{multirow}
\usepackage{upgreek}
\usepackage{booktabs}
\usepackage[verbose]{placeins}
\usepackage{blindtext}
\usepackage{appendix}
\usepackage{enumitem}
\usepackage{color,soul}
\usepackage{rotating}
\usepackage{adjustbox}

\def\code#1{\texttt{#1}}

\newcommand{\wse}{{\it WISE}}

\usepackage{lineno}
\modulolinenumbers[1]
\usepackage{aas_macros}
\usepackage{lscape}

\begin{document}

\title{Optical spectroscopic observations of gamma-ray blazar candidates. X. Results from the 2018--2019 SOAR and OAN-SPM observations of blazar candidates of uncertain type}
%\slugcomment{}
\shorttitle{Optical Campaign X}
\shortauthors{de Menezes et al.}

\author{R. de Menezes \altaffilmark{1,2},
R. A. Amaya-Almaz{\'a}n\altaffilmark{3},
E. J. Marchesini\altaffilmark{2,4,7,8,9},
H. A. Pe\~na-Herazo\altaffilmark{2,3,4,5}, 
F. Massaro\altaffilmark{2,4,5,6},
V. Chavushyan\altaffilmark{3},
A. Paggi\altaffilmark{4,5},
M. Landoni\altaffilmark{10},
N. Masetti\altaffilmark{9,11},
F. Ricci\altaffilmark{12},
R. D'Abrusco\altaffilmark{13},
F. La Franca\altaffilmark{15},
Howard A. Smith\altaffilmark{13},
D. Milisavljevic\altaffilmark{16},
G. Tosti\altaffilmark{17},
E. Jim{\'e}nez-Bail{\'o}n\altaffilmark{18},
C.C. Cheung\altaffilmark{14}
}

\email{raniere.m.menezes@gmail.com}
\altaffiltext{}{R. de Menezes \\ raniere.m.menezes@gmail.com}
\altaffiltext{1}{Universidade de S\~ao Paulo, Departamento de Astronomia, S\~ao Paulo, SP 05508-090, Brazil}
\altaffiltext{2}{Dipartimento di Fisica, Universit\`a degli Studi di Torino, via Pietro Giuria 1, I-10125 Torino, Italy}
\altaffiltext{3}{Instituto Nacional de Astrof\'{i}sica, \'Optica y Electr\'onica, Apartado Postal 51-216, 72000 Puebla, M\'exico}
\altaffiltext{4}{Istituto Nazionale di Fisica Nucleare, Sezione di Torino, I-10125 Torino, Italy}
\altaffiltext{5}{INAF-Osservatorio Astrofisico di Torino, via Osservatorio 20, 10025 Pino Torinese, Italy.}
\altaffiltext{6}{Consorzio Interuniversitario per la Fisica Spaziale (CIFS), via Pietro Giuria 1, I-10125, Torino, Italy.}
\altaffiltext{7}{Facultad de Ciencias Astron\'{o}micas y Geof\'{\i}sicas, Universidad Nacional de La Plata, La Plata, Argentina.}
\altaffiltext{8}{Instituto de Astrof\'{\i}sica de La Plata, CONICET-UNLP, CCT La Plata, La Plata, Argentina.}
\altaffiltext{9}{INAF -- Osservatorio di Astrofisica e Scienza dello Spazio, via Gobetti 93/3, I-40129, Bologna, Italy}
\altaffiltext{10}{INAF-Osservatorio Astronomico di Brera, Via Emilio Bianchi 46, I-23807 Merate, Italy}
\altaffiltext{11}{Departamento de Ciencias F\'isicas, Universidad Andr\'es Bello, Fern\'andez Concha 700, Las Condes, Santiago, Chile}
\altaffiltext{12}{Instituto de Astrof\'{\i}sica and Centro de Astroingenier\'{\i}a, Facultad de F\'{\i}sica, Pontificia Universidad Catolica de Chile, Casilla 306, Santiago 22, Chile}
\altaffiltext{13}{Center for Astrophysics $\mid$ Harvard \& Smithsonian, 60 Garden Street, Cambridge, MA 02138, USA}
\altaffiltext{14}{Naval Research Laboratory, Space Science Division, Code 7650, Washington, DC 20375, USA}
\altaffiltext{15}{Dipartimento di Matematica e Fisica, Universit{\`a} degli Studi Roma Tre, Via della Vasca Navale 84, I-00146, Roma, Italy}
\altaffiltext{16}{Department of Physics and Astronomy, Purdue University, 525 Northwestern Avenue, West Lafayette, IN 47907, USA}
\altaffiltext{17}{Dipartimento di Fisica, Universit{\`a} degli Studi di Perugia,
06123 Perugia, Italy}
\altaffiltext{18}{Instituto de Astronom{\'i}a, Universidad Nacional Aut{\'o}noma de
M{\'e}xico, Apdo. Postal 877, Ensenada, 22800 Baja California, M{\'e}xico}

{\noindent \textbf{Received: 01 Nov 2019 / Accepted: 09 Jan 2020}}

\vspace{0.5cm}

\begin{abstract}

The fourth \textit{Fermi} Large Area Telescope Source Catalog (4FGL) lists over 5000 $\gamma$-ray sources with statistical significance above $4\sigma$. About 23\% of the sources listed in this catalog are unidentified/unassociated $\gamma$-ray sources while $\sim 26\%$ of the sources are classified as blazar candidates of uncertain type (BCUs), lacking optical spectroscopic information. To probe the blazar nature of candidate counterparts of UGSs and BCUs, we started our optical spectroscopic follow up campaign in 2012, which up to date account for more than 350 observed sources. In this paper, the tenth of our campaign, we report on the spectroscopic observations of 37 sources, mostly BCUs, whose observations were carried out predominantly at the Observatorio Astron{\'o}mico Nacional San Pedro M{\'a}rtir and the Southern Astrophysical Research Observatory between August 2018 and September 2019. We confirm the BL Lac nature of 27 sources and the flat spectrum radio quasar nature of three sources. The remaining ones are classified as six BL Lacs galaxy-dominated and one normal galaxy. We were also able to measure the redshifts for 20 sources, including 10 BL Lacs. As in previous analyses, the largest fraction of BCUs revealed to be BL Lac objects.

\end{abstract}


\section{Introduction}
\label{sec:intro}

Blazars are one of the most peculiar and among the rarest class \citep{abdo2010_blazarSED} of active galactic nuclei (AGNs) whose emission, over the whole electromagnetic spectrum, is mainly arising from relativistic particles accelerated in a jet closely aligned to the line of sight \citep{blandford78,urry95}. Together with radio galaxies they constitute the main class of AGNs emitting at MeV-to-TeV energies, with a few exceptions due to a small subset of nearby galaxies \citep{fermi19_4FGL}.

There are two main sub-classes of blazars, distinguished on the basis of their optical spectra: BL Lac objects and flat spectrum radio quasars. The former subclass presents weak emission/absorption lines in their optical spectra with equivalent width (EW) smaller than 5 \AA $ $ \citep{stickel91} or even completely featureless spectra, while sources belonging to the latter show quasar-like optical spectra with broad emission lines.

In the last decade we initiated and carried out an optical spectroscopic campaign to unveil the nature of potential counterparts of unidentified/unassociated $\gamma$-ray sources \citep[UGSs,][]{ugs1,ugs2,massaro2016review} discovered with the \textit{Fermi} Large Area Telescope (LAT) and to confirm the classification of blazar candidates of uncertain type \citep[BCUs;][]{ackermann2015_3LAC}.

Given the continuously increasing number of discovered UGSs by \textit{Fermi}-LAT, our observations are mainly focused on searching blazar-like sources within the positional uncertainty of the UGSs \citep{acero13_UGSs,massaro2015refined,paggi14} and/or eventually BCUs \citep{paper3,crespo16c}. Given the large positional uncertainty of $\gamma$-ray sources, of the order of $\sim$4 arcmin in the third and fourth releases of the \textit{Fermi}-LAT point source catalogs \citep[3FGL and 4FGL,][]{acero15_3FGL,fermi19_4FGL}, potential counterparts, targets of our spectroscopic observations, were selected on the basis of several criteria based, for example, on their mid-infrared colors \citep[MIR;][]{massaro2011_WISEcolors,dabrusco2012_WISEcolors}, presence of X-ray emission \citep{paggi13,takeuchi13} and/or presence of radio sources \citep{schinzel15}.

Here, in the tenth paper of a series devoted to our optical spectroscopic campaign, we focused on the observations of a selected sample of BCUs, that being already associated with \textit{Fermi}-LAT sources, needed an optical spectroscopic classification \citep{ackermann2015_3LAC} to confirm their nature. In the present analysis, for the blazar classification, we adopted the nomenclature of the Roma-BZCAT \citep{massaro15bzcat} distinguishing between BL Lac objects (i.e., BZBs), flat spectrum radio quasars (labelled as BZQs), as well as blazars of galaxy type (BZGs) whereas the spectrum is dominated by the emission from its host galaxy but presents a non-thermal blue continuum likely originated in a relativistic jet (see Section \ref{sec:observations}).

The main goal of our optical spectroscopic campaign is to ``hunt'' blazars among UGSs and BCUs, since they constitute the largest known population of $\gamma$-ray emitters \citep{fermi19_4FGL}. If we confirm a blazar lies within the positional uncertainty region of a UGS, it bolsters its likelihood of being the counterpart of the UGS, due to the scarcity of such sources in the sky. It is thus a valuable information, that has been so far used in every subsequent release of the \textit{Fermi}-LAT catalogs \citep{massaro2015refined}. Even though there has been continued efforts to develop new and alternative methods to increase the association probability \citep[e.g.,][]{dabrusco2019wibrals}, still optical spectroscopy remains an efficient method that provides a precise classification of the potential counterpart of a $\gamma$-ray source.

The state of the art of our campaign can be briefly summarized as follows. We collected spectra for 337 sources to date and classified them as 255 BZBs, 39 BZQs and 20 BZGs, while additional 23 sources have been classified as normal quasars, due to the lack of multifrequency observations, particularly in the radio band, that could confirm the presence of a jet. In particular, 167 sources out of 337 were all BCUs for which the largest fraction (i.e., 114 out of 167, $\sim$70\%) were identified as BZBs, confirming that they are the most elusive subclass of blazars. During our campaign we extensively and continuously searched in the optical databases to exclude targets for which an optical spectrum was already available \citep[see e.g.,][]{massaro14,quest16} and we also compared our results with those present in the literature, due to other spectroscopic campaigns \citep{landoni18,paiano19}, to maintain an updated database. The largest fraction of the sources classified to date come from the archival search in Sloan Digital Sky Survey (SDSS) data, which accounts for a total of 127 spectra; and then more than 100 observations were also carried out in the southern hemisphere with the Southern Astrophysical Research Telescope (SOAR) telescope. All remaining ones were performed with other 2 m- to 6 m-class telescopes as the Kitt Peak National Observatory
(KPNO), the William Herschel Telescope (WHT), Telescopio Nazionale Galileo (TNG), Nordic Optical Telescope (NOT), Observatorio Astrof{\'i}sico Guillermo Haro (OAGH), Observatorio Astron\'omico Nacional San Pedro M{\'a}rtir (OAN-SPM) and Multiple Mirror Telescope (MMT) to name a few.

Our spectroscopic identifications were used by other groups to (i) build the luminosity function of BL Lacs \citep{ajello2014cosmicEvo}; (ii) select potential targets for the Cherenkov Telescope Array \citep[CTA;][]{massaro13,arsioli15}; (iii) obtain stringent limits on the dark matter annihilation in sub-halos \citep[e.g.,][]{zechlin12,berlin14}; (iv) search for counterparts of new flaring $\gamma$-ray sources \citep{bernieri13}; (v) test new $\gamma$-ray detection algorithms \citep{campana15,campana16a}; (vi) perform population studies on the UGSs \citep[e.g.,][]{acero13_UGSs} and (vii) discover the new subclass of radio weak BL Lacs \citep[e.g.,][]{massaro2017radioweak}. All the above studies confirm the legacy value of our results thus motivating the prosecution of our BCU follow up campaign.

The present paper is organized as follows. In \S~\ref{sec:sample} we present the selection criteria for the the BCU sample analysed here, while in \S~\ref{sec:observations} the data reduction procedures for SOAR and OAN-SPM observations are described. In \S~\ref{sec:results} we discuss results obtained, and \S~\ref{sec:summary} is devoted to our summary and conclusions.

\section{Sample selection}
\label{sec:sample}

\begin{figure}
    \centering
    \includegraphics[width=\linewidth]{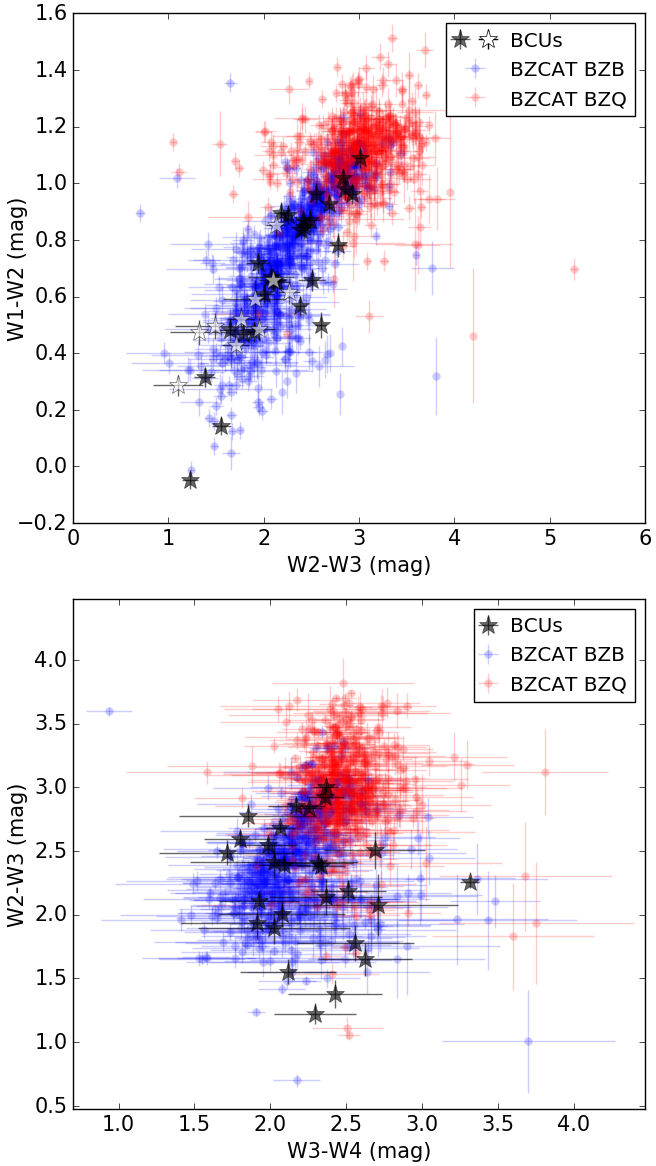}
    \caption{WISE color-color diagrams. Top: distribution of Roma-BZCAT $\gamma$-ray BZBs (blue circles), BZQs (red circles) and our targets (stars) in the W1--W2 $\times$ W2--W3 diagram. Here, black stars represent the targets detected in WISE W4 band, while the white stars lack such a detection. Bottom: same as upper panel but for W2--W3 $\times$ W3--W4.}
    \label{fig:IR_colors}
\end{figure}

Our sample consists of 37 sources predominantly listed as counterparts of $\gamma$-ray sources in the 4FGL catalog (35 sources), as well as in the \textit{Fermi}-LAT 8-year Point Source List\footnote{\url{https://fermi.gsfc.nasa.gov/ssc/data/access/lat/fl8y/}} (FL8Y; 2 sources), a preliminary release of 4FGL based on the same 8-years of data. We highlight that the two sources selected from FL8Y do not appear in 4FGL. The 37 targets are listed as 33 BCUs and 4 BZBs in these $\gamma$-ray catalogs. 

Sources classified as BCUs have unknown blazar nature, as all BCUs i) lack an optical spectra sensitive enough to classify them as BZBs or BZQs, or ii) are classified as blazars of uncertain or transitional types in Roma-BZCAT. However, all BCUs have multiwavelength data indicating a typical blazar-like behaviour \citep[see][]{ackermann2015_3LAC,fermi19_4FGL} and thus, as previously stated, our spectroscopic analysis has been carried out to confirm their real nature. 

In 4FGL, there are 1312 associated sources classified as BCUs (i.e., 26\% of the entire catalog), a fraction significantly higher than on its predecessor 3FGL, for which the BCUs accounted for 19\% of the sample \citep{acero15_3FGL}. All targets selected in our sample have optical magnitude in the $r$ band brighter than 18.5, as listed in the USNO-B catalog \citep{usno2003}. During our observing nights we were able to observe 33 BCUs and 4 sources already classified as BZBs for which we searched for a spectroscopic confirmation and potentially a redshift estimate.

We also verified which of our targets are listed in the latest releases of the Wide-field Infrared Survey Explorer (WISE) Blazar-Like Radio-Loud Sources catalog \citep[WIBRaLS;][]{dabrusco2019wibrals} and the Kernel Density Estimation selected candidate BL Lacs catalog \citep[KDEBLLACS;][]{dabrusco2019wibrals}, both known to present a high number of spectroscopically confirmed blazars \citep{deMenezes2019optical}. We found a total of 19 matches with WIBRaLS (14 classified here as BZBs, 2 BZGs and 3 BZQs; see \S~\ref{sec:results}) and 3 more matches with KDEBLLACS (all classified here as BZBs, see \S~\ref{sec:results}). Additionally, two of our targets, 4FGL J1208.4+6121 and 4FGL J1239.4+0728, are listed in Roma-BZCAT and classified there as blazars of uncertain type.

Although we selected the BCUs mainly from 4FGL, in our previous papers we tend to select sources based on MIR methods, as the \textit{Fermi}-LAT $\gamma$-ray blazars are known to occupy a distinct area in the WISE MIR color-color-color diagram \citep{dabrusco2012_WISEcolors,dabrusco2019wibrals}. Here we compare the distribution of our targets in the WISE MIR color-color diagrams W1-W2 vs. W2-W3 and W2-W3 vs. W3-W4, using the magnitudes measured at the four WISE bands W1[3.4 $\mu$m], W2[4.6 $\mu$m], W3[12 $\mu$m] and W4[22 $\mu$m], with the distribution of MIR colors for $\gamma$-ray blazars listed in Roma-BZCAT, finding that all of them have MIR colors consistent with $\gamma$-ray blazars (Figure \ref{fig:IR_colors}), except perhaps for 4FGL J1612.2+2828/WISE J161217.62+282546.3, the black star in the bottom of both panels, classified as the only normal galaxy in our sample. Among our sources, 27 are detected in WISE W4 band and plotted as black stars in Figure \ref{fig:IR_colors}. The remaining targets are plotted as white stars. The distribution of our targets in these diagrams suggest that their infrared emission is dominated by non-thermal radiation, as expected for $\gamma$-ray blazars.

\wse\ magnitudes used here are in the Vega system and are not corrected for the Galactic extinction. As shown in our previous analyses \citep{ugs1,wibrals}, such correction affect only the magnitude measures at 3.4$\mu$m for sources lying at low Galactic latitudes, and it ranges between 2\% and 5\%, thus being almost negligible.

\section{Observations and data reduction}
\label{sec:observations}

We observed a total of 25 targets with the 2.1 m telescope at OAN-SPM, in Baja California, M{\'e}xico, during several nights in 2018--2019 (see Table \ref{tab:Observation_Log}). The data were collected with the Boller \& Chivens spectrograph, with a slit width of $2.5''$, grating of 300 l/mm, wavelength range from 3800 \AA~ up to 7800 \AA, and resolution of $\sim 14$ \AA.

A single source, 4FGL J0434.7+0922/TXS 0431+092, was observed with the 2.1 m telescope at OAGH in Cananea, M{\'e}xico, using the Boller \& Chivens spectrograph. The collected spectrum has a wavelength range from $\sim 3600$ to 7300 \AA~and was acquired with a slit width of $2.5''$ and grating of 150 l/mm, having a final resolution of $\sim 14$ \AA.

The data for the remaining 11 targets were collected with the 4.1 m SOAR telescope at Cerro Pach{\'o}n, Chile, in remote observing mode on May 22 and 23 of 2019. We used the single, long slit mode of the Goodman High Throughput Spectrograph \citep{clemens04} with slit width of 1$''$ and a grating of 400 l/mm, giving a dispersion of $\sim 3$\AA~ per pixel in a spectral range from $\sim 4100$ \AA~up to 7900 \AA, and resolution of $\sim 6$\AA. 

In Table \ref{tab:Observation_Log} we report the log of all observations, giving the name of the $\gamma$-ray source and its associated counterpart in the \textit{Fermi}-LAT catalogs, the associated counterpart in WISE, the telescope used, the average signal to noise ratio (SNR), the exposure time and the date of observation.

The data were reduced using standard procedures, with the images trimmed, bias subtracted, corrected for flat field and stacked using standard \code{IRAF} packages \citep{tody86}. The cosmic ray removal was performed with Python packages \code{ccdproc} \citep{craig2015ccdproc} and \code{astropy} \citep{2013A&Aastropy1,2018AJastropy2}, in the case of SOAR data, and using the L.A. Cosmic IRAF algorithm (\citeauthor{vandockum2001lacosmic} \citeyear{vandockum2001lacosmic}) in the case of OAN-SPM data. All spectra were then flux and wavelength calibrated using, respectively, a standard spectrophotometric star and a Hg-Ar (SOAR) or CuNeHeAr (OAN-SPM) lamp. The corrections for Galactic extinction were made using the reddening law of \citeauthor{cardelli89} \citeyearpar{cardelli89} and values of visual extinction $A_{v}$ computed based on \citeauthor{schlafly11}~\citeyearpar{schlafly11} assuming a visual extinction to reddening ratio $A_v/E(B-V) = 3.1$. Finally, we performed a box smoothing for visual representation, normalized the spectra using a 5th degree polynomial fitted to the continuum to highlight possible spectroscopic features, and visually identified the emission/absorption lines (Figure \ref{fig:results}).

\section{Analysis and results}
\label{sec:results}

Thanks to the new optical spectra collected, we are able to confirm the blazar nature for 36 targets in our sample, consisting of 27 BZBs, 6 BZGs, and 3 BZQs. Additionally, we identified a normal galaxy listed in 4FGL as the counterpart of 4FGL J1612.2+2828. 

We obtained the redshifts ($z$) for 20 of the targets, including 10 BZBs, all with relatively low $z$ distributed in the range 0.12--0.28, as expected for BZBs. Particularly we successfully measured redshifts for 3 of the 4 BZBs listed in the 4FGL (see Section \ref{sec:sample}). They are 4FGL J0420.2+4012/WISE J042013.43+401121.2 with z = 0.132, 4FGL J2220.5+2813/WISE J222028.72+281355.8 with $z = 0.148$ and 4FGL J2358.3+3830/WISE J235825.19+382856.5 with $z = 0.200$. For the last source, we observed emission lines with EW $> 5$ \AA, and thus classified it as a BZQ.

All results achieved for the entire sample are summarized in Table \ref{tab:results}, where we show the name of the $\gamma$-ray sources as listed in the \textit{Fermi}-LAT catalogs, the name of the WISE counterparts, the redshifts when available, the spectral classes and information about the observed absorption/emission lines.

The optical spectra and finding charts for the analysed sources are available in Appendix \ref{sec.foo}, ordered by right ascension. Finding charts were obtained from ESO Online Digitized Sky Survey\footnote{\url{http://archive.eso.org/dss/dss}} and have a size of $5' \times 5'$. The name of each emission/absorption line is also indicated in the figure. Telluric absorption lines are labelled with $\bigoplus$. 

Criteria adopted to distinguishing between BZBs, BZGs and normal galaxies are based on the observed Ca II H\&K break contrast. We computed the ratio $C = (F_+ - F_-)/F_+$ where $F_+$ and $F_-$ are the mean flux densities measured in small ranges of 200 \AA~ just above (4050-4250 \AA) and below (3750-3950 \AA) the Ca II H\&K break, respectively. A value of $C < 0.4$ means that the source has a blue continuum emission dominated by non-thermal radiation likely originated in a relativistic jet \citep{marcha1996CaBreak,landt2002CaBreak,massaro12b}. Then sources with $C > 0.4$ are classified as normal galaxies, sources with $0.25 < C \leq 0.4$ are classified as BZG, while values of $C \leq 0.25$ lead towards a BZB or BZQ classification.

To distinguish between BZBs and BZQs we computed the rest frame equivalent width (EW) of their emission lines (when available) and classified them as BZBs if all lines had EW $\leq 5$\AA. In particular, we found only 3 BCUs having emission lines with EW $> 5$\AA, leading to a BZQ classification. 

For those sources for which a redshift estimate has been provided, the redshift uncertainties are of the order of $\Delta$z = 0.001 for the data collected with SOAR, while those pointed with OAGH and OAN-SPM have $\Delta$z $\sim$ 0.004.

\section{Summary and Conclusions}
\label{sec:summary}

We presented the tenth paper of our optical spectroscopic campaign aiming at the identification of BCUs listed in \textit{Fermi}-LAT catalogs. We collected and analysed data for a total of 37 sources, confirming the blazar nature for 36 of them. Our results are summarized as follows:

\begin{itemize}
    \item Among the 33 sources classified as BCUs in the \textit{Fermi}-LAT catalogs, 24 of them (73\%) are found to be BZBs, 6 are BZGs (18\%), 2 are BZQs (6\%) and 1 is a normal galaxy (3\%). We also measured the redshifts for 17 of them.
    \item We observed 4 sources classified as BZBs in 4FGL and were able to measure the redshifts for 3 of them: 4FGL J0420.2+4012, 4FGL J2220.5+2813 and 4FGL J2358.3+3830, with z = 0.132, z = 0.148 and z = 0.200 respectively. In particular, 4FGL J2358.3+3830 presents emission lines with EW $> 5$\AA, thus being classified as BZQ.
    \item All sources classified as blazars (36 out of 37) have MIR colors consistent with what is expected for $\gamma$-ray blazars (Figure \ref{fig:IR_colors}), except perhaps for the counterpart of 4FGL J1608.0-2038.
\end{itemize}

The results here, as in the previous papers of our campaign \citep[see e.g.,][]{marchesini19,pena2019Ap&SS_optCampIX}, confirm that the largest fraction of BCUs tend to be BZBs. These objects, although very common in the $\gamma$-ray sky, continue to be one of the most elusive classes of $\gamma$-ray sources.

Our optical spectroscopic campaign is still ongoing with observing nights scheduled for the current and next semesters. New results will be made available as soon as they are collected and analysed. The next paper of the campaign will focus mainly on UGSs (Pe{\~n}a-Herazo et al. 2019) observed with the 4 m Blanco telescope at Cerro Tololo, Chile. We observed a total of 374 targets to date, mostly BCUs and UGSs, but including a small fraction (i.e., of the order of a few percent) of known blazars, especially BL Lac objects, trying to catch them in a low optical state and obtain a $z$ estimate. Particularly, out of 200 sources previously classified as BCUs, we found that 141 of them are BZBs.

\begin{sidewaystable*}
    \centering
    \begin{adjustbox}{scale=0.85,center}
     \begin{tabular}{|l|l|l|c|c|c|r|c|c|}
\hline
  \multicolumn{1}{|c|}{\textit{Fermi}-LAT name} &
  \multicolumn{1}{c|}{Association} &
  \multicolumn{1}{c|}{WISE name} &
  \multicolumn{1}{c|}{R.A. (J2000)} &
  \multicolumn{1}{c|}{Dec. (J2000)} &
  \multicolumn{1}{c|}{Telescope} &
  \multicolumn{1}{c|}{SNR} &
  \multicolumn{1}{c|}{Exposure} &
  \multicolumn{1}{c|}{Obs. date} \\
    &   &   & (hh:mm:ss) & (dd:mm:ss) &   &   & (s) & (dd/mm/yyyy) \\
  \hline
  4FGL J0015.9+2440 & GB6 J0016+2440 & J001603.62+244014.7 & 00:16:04 & +24:40:15 & OAN & 15 & 1800 & 17/10/2018\\
  4FGL J0119.6+4158 & 2MASX J01200274+4200139 & J012002.76+420013.8 & 01:20:03 & +42:00:14 & OAN & 11 & 1800 & 17/10/2018\\
  4FGL J0204.3+2417 & B2 0201+24 & J020421.54+241750.7 & 02:04:22 & +24:17:51 & OAN & 8 & 1800 & 18/10/2018\\
  4FGL J0325.3+3332 & 2MASX J03251760+3332435 & J032517.58+333243.7 & 03:25:18 & +33:32:44 & OAN & 9 & 1200 & 17/10/2018\\
  4FGL J0420.2+4012 & 1RXS J042008.3+401138 & J042013.43+401121.2 & 04:20:13 & +40:11:21 & OAN & 10 & 1800 & 19/10/2018\\
  4FGL J0434.7+0922 & TXS 0431+092 & J043440.98+092348.6 & 04:34:41 & +09:23:49 & OAGH & 5 & 1800 & 08/11/2018\\
  4FGL J0442.7+6142 & GB6 J0442+6140 & J044240.66+614039.4* & 04:42:41 & +61:40:39 & OAN & 5 & 1800 & 17/10/2018\\
  FL8Y J0516.6+2742 & VCS J0516+2743 & J051640.48+274310.2 & 05:16:40 & +27:43:10 & OAN & 8 & 3600 & 18/10/2018\\
  4FGL J0602.7-0007 & PMN J0602-0004 & J060242.88-000426.7 & 06:02:43 & -00:04:27 & OAN & 9 & 3600 & 17/10/2018\\
  4FGL J0842.7+6656 & TXS 0838+671 & J084243.19+665729.4 & 08:42:43 & +66:57:29 & OAN & 10 & 1800 & 07/04/2019\\
  4FGL J0928.7-3529 & NVSS J092849-352947 & J092849.82-352948.8 & 09:28:50 & -35:29:49 & SOAR & 16 & 2 x 1200 & 22/05/2019\\
  4FGL J1028.3+3108 & TXS 1025+313 & J102817.61+310734.4 & 10:28:18 & +31:07:34 & OAN & 6 & 3600 & 06/04/2019\\
  4FGL J1042.1-4128 & 1RXS J104204.1-412936 & J104203.00-412929.9 & 10:42:03 & -41:29:30 & SOAR & 20 & 2 x 1500 & 23/05/2019\\
  4FGL J1156.6-2248 & NVSS J115633-225004 & J115633.22-225004.2 & 11:56:33 & -22:50:04 & SOAR & 20 & 2 x 1300 & 23/05/2019\\
  4FGL J1208.4+6121 & RGB J1208+613 & J120837.13+612106.4 & 12:08:37 & +61:21:06 & OAN & 9 & 1800 & 07/04/2019\\
  4FGL J1213.8-4345 & PMN J1213-4343 & J121350.38-434324.6 & 12:13:50 & -43:43:25 & SOAR & 16 & 2 x 1500 & 23/05/2019\\
  4FGL J1239.4+0728 & PKS 1236+077 & J123924.58+073017.2 & 12:39:25 & +07:30:17 & OAN & 17 & 1800 & 08/04/2019\\
  4FGL J1310.6+2449 & MG2 J131037+2447 & J131038.49+244822.7 & 13:10:38 & +24:48:23 & OAN & 7 & 1800 & 08/04/2019\\
  4FGL J1351.4-1529 & 2MASX J13511746-1530155 & J135117.45-153016.0 & 13:51:17 & -15:30:16 & SOAR & 19 & 2 x 1300 & 23/05/2019\\
  4FGL J1400.2-4010 & 2MASS J14002208-4008235 & J140022.08-400823.6 & 14:00:22 & -40:08:24 & SOAR & 12 & 2 x 1800 & 22/05/2019\\
  4FGL J1440.0-2343 & PMN J1439-2341 & J143959.46-234141.0 & 14:39:59 & -23:41:41 & SOAR & 19 & 2 x 1500 & 22/05/2019\\
  4FGL J1546.1-1003 & PMN J1546-1003 & J154611.48-100326.1 & 15:46:11 & -10:03:26 & SOAR & 21 & 2 x 1200 & 23/05/2019\\
  4FGL J1550.8-0822 & NVSS J155053-082247 & J155053.26-082246.7 & 15:50:53 & -08:22:47 & OAN & 8 & 1800 & 05/08/2018\\
  4FGL J1606.9+5919 & 1RXS J160709.7+592115 & J160708.14+592129.5 & 16:07:08 & +59:21:30 & OAN & 10 & 1800 & 05/06/2019\\
  4FGL J1608.0-2038 & NVSS J160756-203942 & J160756.90-203943.5 & 16:07:57 & -20:39:44 & SOAR & 18 & 2 x 1700 & 22/05/2019\\
  4FGL J1612.2+2828 & TXS 1610+285 & J161217.62+282546.3 & 16:12:18 & +28:25:46 & OAN & 11 & 1200 & 06/06/2019\\
  4FGL J1707.5+1649 & MG1 J170732+1649 & J170731.55+164844.6 & 17:07:32 & +16:48:45 & OAN & 8 & 1800 & 07/08/2018\\
  4FGL J1741.2+5739 & NVSS J174111+573812 & J174111.70+573812.4 & 17:41:12 & +57:38:12 & OAN & 6 & 1800 & 04/07/2019\\
  4FGL J1808.2+3500 & MG2 J180813+3501 & J180811.52+350118.7 & 18:08:12 & +35:01:19 & OAN & 14 & 1200 & 08/04/2019\\
  4FGL J1830.0-5225 & SUMSS J183004-522618 & J183004.32-522618.8 & 18:30:04 & -52:26:19 & SOAR & 13 & 2 x 1600 & 23/05/2019\\
  4FGL J1831.9+3820 & 1RXS J183202.2+382132 & J183200.98+382137.0 & 18:32:01 & +38:21:37 & OAN & 8 & 3600 & 07/08/2018\\
  4FGL J1838.0-5959 & SUMSS J183806-600033 & J183806.74-600032.1 & 18:38:07 & -60:00:32 & SOAR & 19 & 2 x 1200 & 23/05/2019\\
  4FGL J1858.7+5708 & 87GB 185759.9+570427 & J185853.50+570809.5 & 18:58:54 & +57:08:10 & OAN & 7 & 3600 & 08/08/2018\\
  FL8Y J2201.6+2953 & 2WHSP J220123.8+294934 & J220123.82+294934.5 & 22:01:24 & +29:49:35 & OAN & 10 & 1800 & 05/08/2018\\
  4FGL J2220.5+2813 & RX J2220.4+2814 & J222028.72+281355.8 & 22:20:29 & +28:13:56 & OAN & 11 & 1800 & 03/08/2018\\
  4FGL J2236.6+3706 & NVSS J223626+370713 & J223626.34+370713.5 & 22:36:26 & +37:07:14 & OAN & 6 & 1800 & 18/10/2018\\
  4FGL J2358.3+3830 & B3 2355+382 & J235825.19+382856.5 & 23:58:25 & +38:28:57 & OAN & 7 & 3600 & 04/08/2018\\
\hline\end{tabular}
\end{adjustbox}
    \caption{Observation log. Col. (1) The name reported in the \textit{Fermi}-LAT catalogs (i.e., FL8Y and 4FGL); Col. (2) the name of the associated counterpart listed in the \textit{Fermi}-LAT catalogs; Col. (3) the WISE name of the associated counterpart; Col. (4) and Col. (5) The coordinates (Equinox J2000) of the pointed source; Col. (6) the telescope used to carry out the observation; Col. (7) The average signal to noise ratio; Col. (8) the exposure time and Col. (9) the observation date. The source tagged with ``*" is found only in the AllWISE Multiepoch Photometry Table.
    }
    \label{tab:Observation_Log}
\end{sidewaystable*}

% r magnitudes as listed in the United States Naval Observatory B1.0 Catalog

\begin{table*}
    \centering
        \begin{adjustbox}{scale=0.85,center}
    \begin{tabular}{|l|l|l|l|l|l|l|l|l|}
\hline
  \multicolumn{1}{|c|}{\textit{Fermi}-LAT name} &
  \multicolumn{1}{c|}{WISE counterpart} &
  \multicolumn{1}{c|}{Class} &
  \multicolumn{1}{c|}{C} &
  \multicolumn{1}{c|}{$z$} &
  \multicolumn{1}{c|}{Line} &
  \multicolumn{1}{c|}{EW (\AA)} &
  \multicolumn{1}{c|}{$\lambda_{obs}$ (\AA)} &
  \multicolumn{1}{c|}{Type} \\
\hline
  4FGL J0015.9+2440 & J001603.62+244014.7 & BZB &  &  &  &  &  & \\
  4FGL J0119.6+4158 & J012002.76+420013.8 & BZG & 0.30 & 0.109 & [O II] & 5 & 4132 & E\\
   &  &  &  &  & Ca II H\&K & 10; 9 & 4362; 4399 & A\\
   &  &  &  &  & G band & 5 & 4773 & A\\
   &  &  &  &  & Mg I & 5 & 5737 & A\\
   &  &  &  &  & Na I & 5 & 6532 & A\\
  4FGL J0204.3+2417 & J020421.54+241750.7 & BZB & 0.21 & 0.210 & Ca II H\&K & 7; 5 & 4760; 4803 & A\\
   &  &  &  &  & G band & 3 & 5510 & A\\
   &  &  &  &  & Mg I & 5 & 6267 & A\\
   &  &  &  &  & Na I & 4 & 7131 & A\\
  4FGL J0325.3+3332 & J032517.58+333243.7 & BZG & 0.36 & 0.128 & Ca II H\&K & 12; 9 & 4440; 4477 & A\\
   &  &  &  &  & G band & 6 & 4855 & A\\
   &  &  &  &  & Mg I & 6 & 5838 & A\\
  4FGL J0420.2+4012 & J042013.43+401121.2 & BZB & 0.01 & 0.132 & G band & 4 & 4874 & A\\
   &  &  &  &  & Mg I & 6 & 5860 & A\\
   &  &  &  &  & Na I & 4 & 6676 & A\\
  4FGL J0434.7+0922 & J043440.98+092348.6 & BZB &  &  &  &  &  & \\
  4FGL J0442.7+6142 & J044240.66+614039.4 & BZB &  &  &  &  &  & \\
  FL8Y J0516.6+2742 & J051640.48+274310.2 & BZG & 0.34 & 0.060 & [O II] & 5 & 3952 & E\\
   &  &  &  &  & Ca II H\&K & 8; 6 & 4172; 4206 & A\\
   &  &  &  &  & G band & 3 & 4566 & A\\
   &  &  &  &  & Mg I & 5 & 5492 & A\\
   &  &  &  &  & Na I & 4 & 6251 & A\\
   &  &  &  &  & H$_{\alpha}$ & 5 & 6962 & E\\
   &  &  &  &  & [N II] & 7 & 6984 & E\\
   &  &  &  &  & [S II] & 2 & 7125 & E\\
   &  &  &  &  & [S II] & 2 & 7141 & E\\
  4FGL J0602.7-0007 & J060242.88-000426.7 & BZG & 0.29 & 0.118 & Ca II H\&K & 5 & 4447 & A\\
   &  &  &  &  & G band & 6 & 4810 & A\\
   &  &  &  &  & Mg I & 6 & 5788 & A\\
  4FGL J0842.7+6656 & J084243.19+665729.4 & BZB & 0.24 & 0.121 & Ca II H\&K & 8; 8 & 4412; 4450 & A\\
   &  &  &  &  & G band & 3 & 4829 & A\\
   &  &  &  &  & Mg I & 4 & 5803 & A\\
   &  &  &  &  & Na I & 3 & 6609 & A\\
  4FGL J0928.7-3529 & J092849.82-352948.8 & BZB &  &  &  &  &  & \\
  4FGL J1028.3+3108 & J102817.61+310734.4 & BZB &  &  &  &  &  & \\
  4FGL J1042.1-4128 & J104203.00-412929.9 & BZB &  &  &  &  &  & \\
  4FGL J1156.6-2248 & J115633.22-225004.2 & BZB &  & 0.860? & Mg II? & 2; 2 & 5200; 5214 & A\\
  4FGL J1208.4+6121 & J120837.13+612106.4 & BZB & 0.13 & 0.275 & Ca II H\&K & 3; 3 & 5018; 5062 & A\\
   &  &  &  &  & G band & 4 & 5486 & A\\
   &  &  &  &  & Mg I & 4 & 6604 & A\\
  4FGL J1213.8-4345 & J121350.38-434324.6 & BZB &  &  &  &  &  & \\
  4FGL J1239.4+0728 & J123924.58+073017.2 & BZB &  &  &  &  &  & \\
  4FGL J1310.6+2449 & J131038.49+244822.7 & BZG & 0.29 & 0.226 & Ca II H\&K & 9; 10 & 4825; 4866 & A\\
   &  &  &  &  & G band & 5 & 5269 & A\\
   &  &  &  &  & Mg I & 4 & 6342 & A\\
\hline\end{tabular}
\end{adjustbox}
    \caption{Summary of our spectroscopic identifications. Col. (1) the name reported in the \textit{Fermi}-LAT catalogs (i.e., FL8Y and 4FGL); Col. (2) the WISE name of the associated counterpart; Col. (3) the source class (i.e., BZB for BL Lac objects, BZQ for Blazars of quasar type, BZG for blazars of galaxy type); Col. (4) the measured Ca II H\&K contrast; Col. (5) measured redshift; Col. (6) Emission or Absorption lines identified; Col. (7) Measured equivalent width (EW); Col. (8) Observed wavelength of each emission/absorption line identified; Col. (9) type of line: absorption (A) or emission (E).}
    \label{tab:results}
\end{table*}

\begin{table*}
%\ContinuedFloat
    \centering
    \begin{adjustbox}{scale=0.85,center}
    \begin{tabular}{|l|l|l|l|l|l|l|l|l|}
    
\hline
  \multicolumn{1}{|c|}{\textit{Fermi}-LAT name} &
  \multicolumn{1}{c|}{WISE counterpart} &
  \multicolumn{1}{c|}{Class} &
  \multicolumn{1}{c|}{C} &
  \multicolumn{1}{c|}{$z$} &
  \multicolumn{1}{c|}{Line} &
  \multicolumn{1}{c|}{EW (\AA)} &
  \multicolumn{1}{c|}{$\lambda_{obs}$ (\AA)} &
  \multicolumn{1}{c|}{Type} \\
\hline  
  4FGL J1351.4-1529 & J135117.45-153016.0 & BZB & 0.01 & 0.285 & Ca II H\&K & 1; 2 & 5061; 5101 & A\\
   &  &  &  &  & G band & 2 & 5530 & A\\
   &  &  &  &  & Mg I & 1 & 6656 & A\\
  4FGL J1400.2-4010 & J140022.08-400823.6 & BZB & 0.24 & 0.203 & Ca II H\&K & 7 & 4775 & A\\
   &  &  &  &  & G band & 4 & 5177 & A\\
   &  &  &  &  & Mg I & 5 & 6226 & A\\
   &  &  &  &  & Na I & 3 & 7090 & A\\
  4FGL J1440.0-2343 & J143959.46-234141.0 & BZB &  &  &  &  &  & \\
  4FGL J1546.1-1003 & J154611.48-100326.1 & BZB &  &  &  &  &  & \\
  4FGL J1550.8-0822 & J155053.26-082246.7 & BZB &  &  &  &  &  & \\
  4FGL J1606.9+5919 & J160708.14+592129.5 & BZB & 0.18 & 0.132 & [O II] & 4 & 4217 & E\\
   &  &  &  &  & Ca II H\&K & 4; 4 & 4453; 4491 & A\\
   &  &  &  &  & G band & 3 & 4872 & A\\
   &  &  &  &  & Mg I & 3 & 5855 & A\\
   &  &  &  &  & Na I & 2 & 6667 & A\\
  4FGL J1608.0-2038 & J160756.90-203943.5 & BZB &  &  &  &  &  & \\
  4FGL J1612.2+2828 & J161217.62+282546.3 & galaxy & 0.47 & 0.053 & Ca II H\&K & 10; 9 & 4148; 4183 & A\\
   &  &  &  &  & G band & 7 & 4537 & A\\
   &  &  &  &  & Mg I & 5 & 5454 & A\\
   &  &  &  &  & Na I & 4 & 6211 & A\\
   &  &  &  &  & H$_{\alpha}$ & 2 & 6915 & E\\
   &  &  &  &  & [N II] & 2 & 6938 & E\\
   &  &  &  &  & [S II]? & 9 & 7081 & E\\
  4FGL J1707.5+1649 & J170731.55+164844.6 & BZQ & 0.10 & 0.291 & [O II] & 7 & 4817 & E\\
   &  &  &  &  & Ca II H\&K & 1 & 5133 & A\\
   &  &  &  &  & G band & 1 & 5558 & A\\
   &  &  &  &  & [O III] & 3; 5 & 6409; 6470 & E\\
  4FGL J1741.2+5739 & J174111.70+573812.4 & BZB &  &  &  &  &  & \\
  4FGL J1808.2+3500 & J180811.52+350118.7 & BZB &  & 0.269? & [O II]? & 5 & 4721 & E\\
  4FGL J1830.0-5225 & J183004.32-522618.8 & BZB &  &  &  &  &  & \\
  4FGL J1831.9+3820 & J183200.98+382137.0 & BZG & 0.30 & 0.216 & Ca II H\&K & 8; 7 & 4785; 4826 & A\\
   &  &  &  &  & G band & 4 & 5234 & A\\
   &  &  &  &  & Mg I & 2 & 6304 & A\\
   &  &  &  &  & Na I & 4 & 7168 & A\\
  4FGL J1838.0-5959 & J183806.74-600032.1 & BZB &  &  &  &  &  & \\
  4FGL J1858.7+5708 & J185853.50+570809.5 & BZQ & -0.02 & 0.076 & [O II] & 4 & 4017 & E\\
   &  &  &  &  & H$_{\beta}$ & 2 & 5236 & E\\
   &  &  &  &  & [O III] & 4;5 & 5346; 5393 & E\\
   &  &  &  &  & H$_{\alpha}$ & 12 & 7067 & E\\
  FL8Y J2201.6+2953 & J220123.82+294934.5 & BZB & 0.22 & 0.149 & Ca II H\&K & 7; 4 & 4521; 4560 & A\\
   &  &  &  &  & G band & 3 & 4945 & A\\
   &  &  &  &  & Mg I & 3 & 5951 & A\\ 
  4FGL J2220.5+2813 & J222028.72+281355.8 & BZB & 0.24 & 0.148 & Ca II H\&K & 8; 5 & 4519; 4558 & A\\
   &  &  &  &  & G band & 4 & 4944 & A\\
   &  &  &  &  & Mg I & 4 & 5945 & A\\
   &  &  &  &  & Na I & 4 & 6768 & A\\
  4FGL J2236.6+3706 & J223626.34+370713.5 & BZB & 0.15 & 0.235 & Ca II H\&K & 6; 5 & 4861; 4899 & A\\
   &  &  &  &  & G band & 3 & 5318 & A\\
   &  &  &  &  & Mg I & 4 & 6393 & A\\
  4FGL J2358.3+3830 & J235825.19+382856.5 & BZQ & 0.00 & 0.200 & [O II] & 9 & 4474 & E\\
   &  &  &  &  & H$_{\beta}$ & 1 & 5837 & E\\
   &  &  &  &  & [O III] & 4; 10 & 5953; 6011 & E\\
   &  &  &  &  & [Fe VII]? & 4 & 6874 & A\\
\hline\end{tabular}
\end{adjustbox}
    \caption{Continued from Table 2.}
\end{table*}

\acknowledgements

This work was supported by FAPESP (Funda\c{c}\~ao de Amparo \`a Pesquisa do Estado de S\~ao Paulo) under grants 2016/25484-9 and 2018/24801-6 (R.M.). R.A.A-A. and H.P. acknowledge support from the CONACyT program for Ph.D. studies. The work of F.M. and A.P. is partially supported by the ``Departments of Excellence 2018-2022'' Grant awarded by the Italian Ministry of Education, University and Research (MIUR) (L. 232/2016) and made use of resources provided by the Compagnia di San Paolo for the grant awarded on the BLENV project (S1618\_L1\_MASF\_01) and by the Ministry of Education, Universities and Research for the grant MASF\_FFABR\_17\_01. F.M. also acknowledges financial contribution from the agreement ASI-INAF n.2017-14-H.0 while A.P. the financial support from the Consorzio Interuniversitario per la Fisica Spaziale (CIFS) under the agreement related to the grant MASF\_CONTR\_FIN\_18\_02. F.R. acknowledges support from FONDECYT Postdoctorado 3180506 and CONICYT project Basal AFB-170002. C.C.C. at NRL was supported by the Chief of Naval Research.

This research made use of Astropy,\footnote{\url{http://www.astropy.org}} a community-developed core Python package for Astronomy \citep{2013A&Aastropy1,2018AJastropy2}, ccdproc, an Astropy package for image reduction \citep{craig2015ccdproc}, and TOPCAT\footnote{\url{http://www.star.bris.ac.uk/~mbt/topcat/}} \citep{taylor05} for the preparation and manipulation of the tabular data.

%SOAR
This project makes use of data products from the Southern Astrophysical Research (SOAR) telescope, which is a joint project of the Minist\'{e}rio da Ci\^{e}ncia, Tecnologia, Inova\c{c}\~{o}es e Comunica\c{c}\~{o}es (MCTIC) da Rep\'{u}blica Federativa do Brasil, the U.S. National Optical Astronomy Observatory (NOAO), the University of North Carolina at Chapel Hill (UNC), and Michigan State University (MSU).

%OAN
This project makes use of spectroscopic observations acquired at the 2.1 m telescope of the Observatorio Astron{\'o}mico Nacional San Pedro M{\'a}rtir (OAN-SPM), Baja California, M{\'e}xico. We also thank the staff at the Observatorio Astrof{\'i}sico Guillermo Haro (OAGH) for all their help during the observation runs.

% WISE
This publication makes use of data products from the Wide-field Infrared Survey Explorer, 
which is a joint project of the University of California, Los Angeles, and 
the Jet Propulsion Laboratory/California Institute of Technology, 
funded by the National Aeronautics and Space Administration.

\bibliographystyle{apj}

\appendix\section{Optical spectra and finding charts}\label{sec.foo}

Here we present all the spectra collected together with their respective finding charts extracted from ESO Online Digitized Sky Survey (Figure \ref{fig:results}).

\begin{figure*}
    \centering
    \includegraphics[scale=0.5]{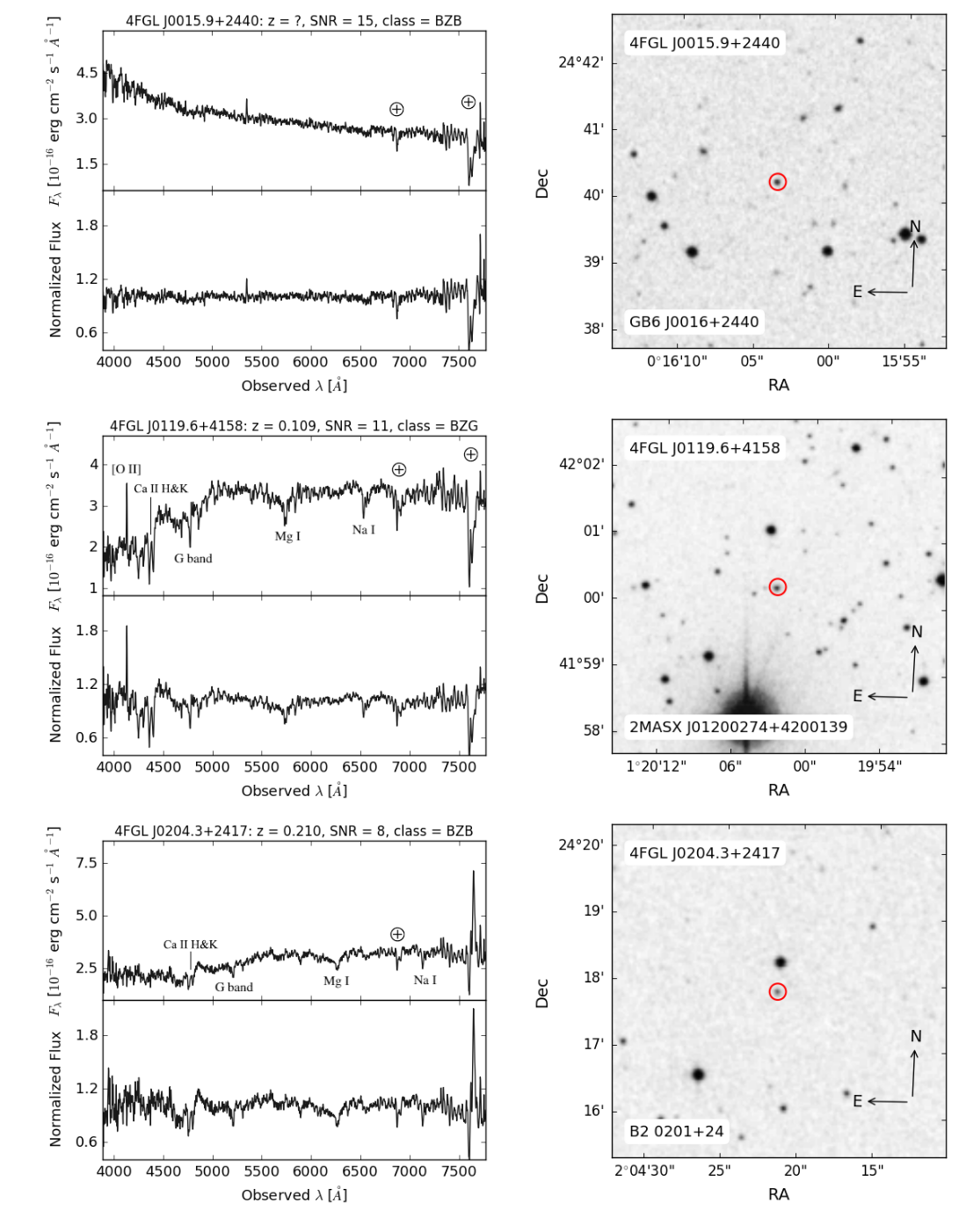}
    \caption{Left: flux calibrated and normalized optical spectra. Right: finding charts. The name of the $\gamma$-ray sources and their respective counterparts listed in the \textit{Fermi}-LAT catalogs are indicated, together with the spectral class and average SNR. Redshifts and emission/absorption lines are shown whenever possible.}
    \label{fig:results}
\end{figure*}

\begin{figure*}
%\ContinuedFloat
    \centering
    \includegraphics[scale=0.5]{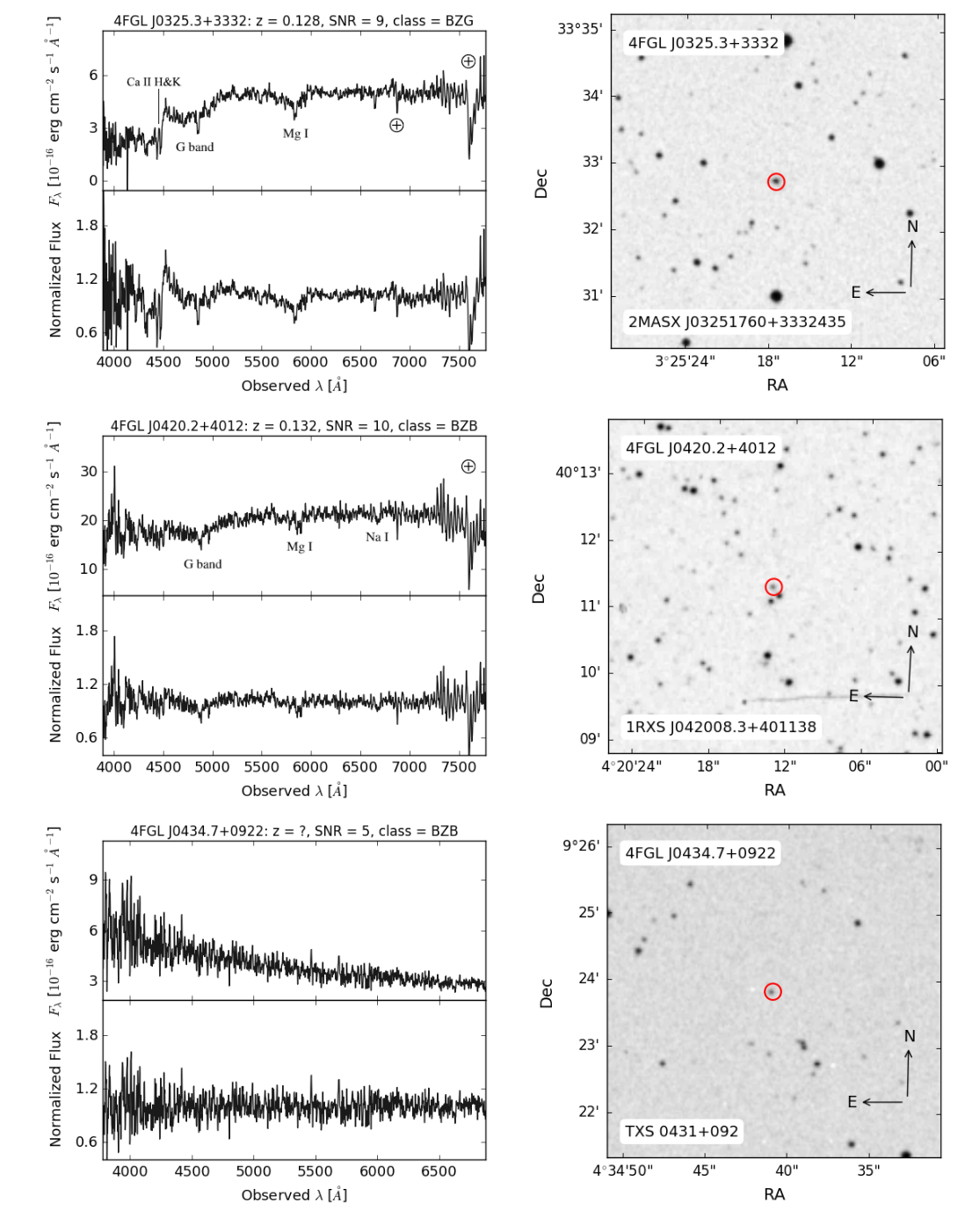}
    \caption{Continued from Figure 2.}
\end{figure*}

\begin{figure*}
%\ContinuedFloat
    \centering
    \includegraphics[scale=0.5]{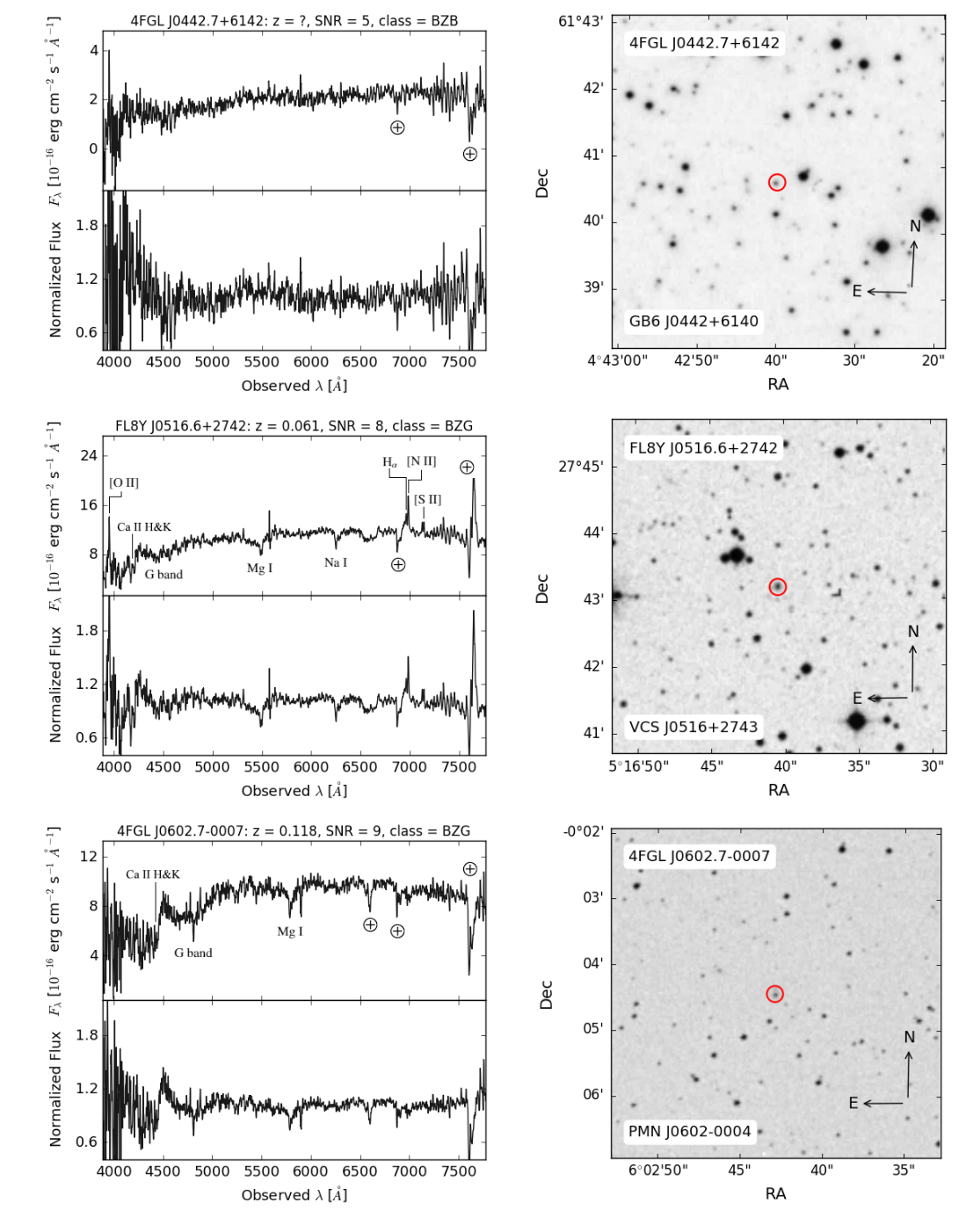}
    \caption{Continued from Figure 2.}
\end{figure*}

\begin{figure*}
%\ContinuedFloat
    \centering
    \includegraphics[scale=0.5]{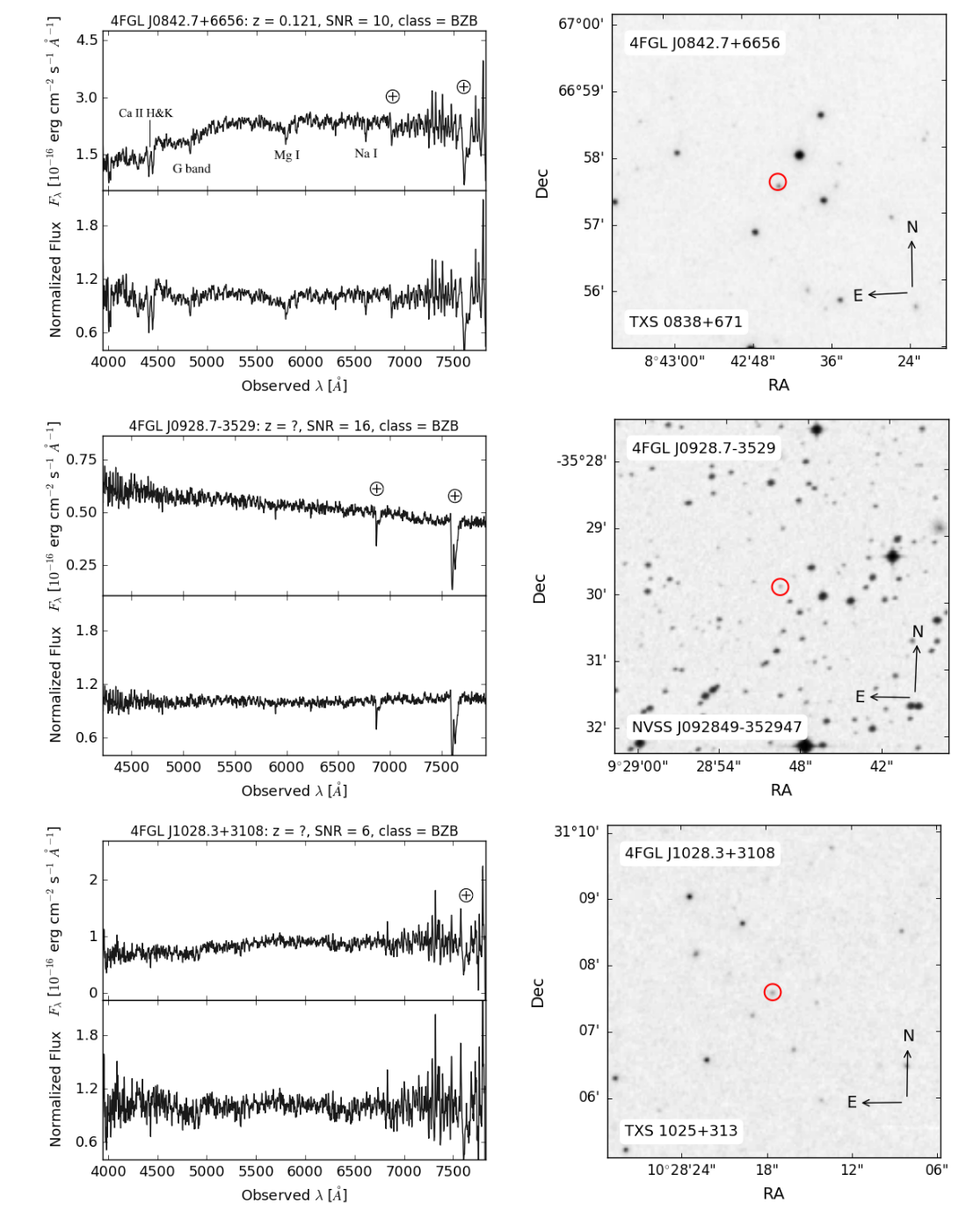}
    \caption{Continued from Figure 2.}
\end{figure*}

\begin{figure*}
%\ContinuedFloat
    \centering
    \includegraphics[scale=0.5]{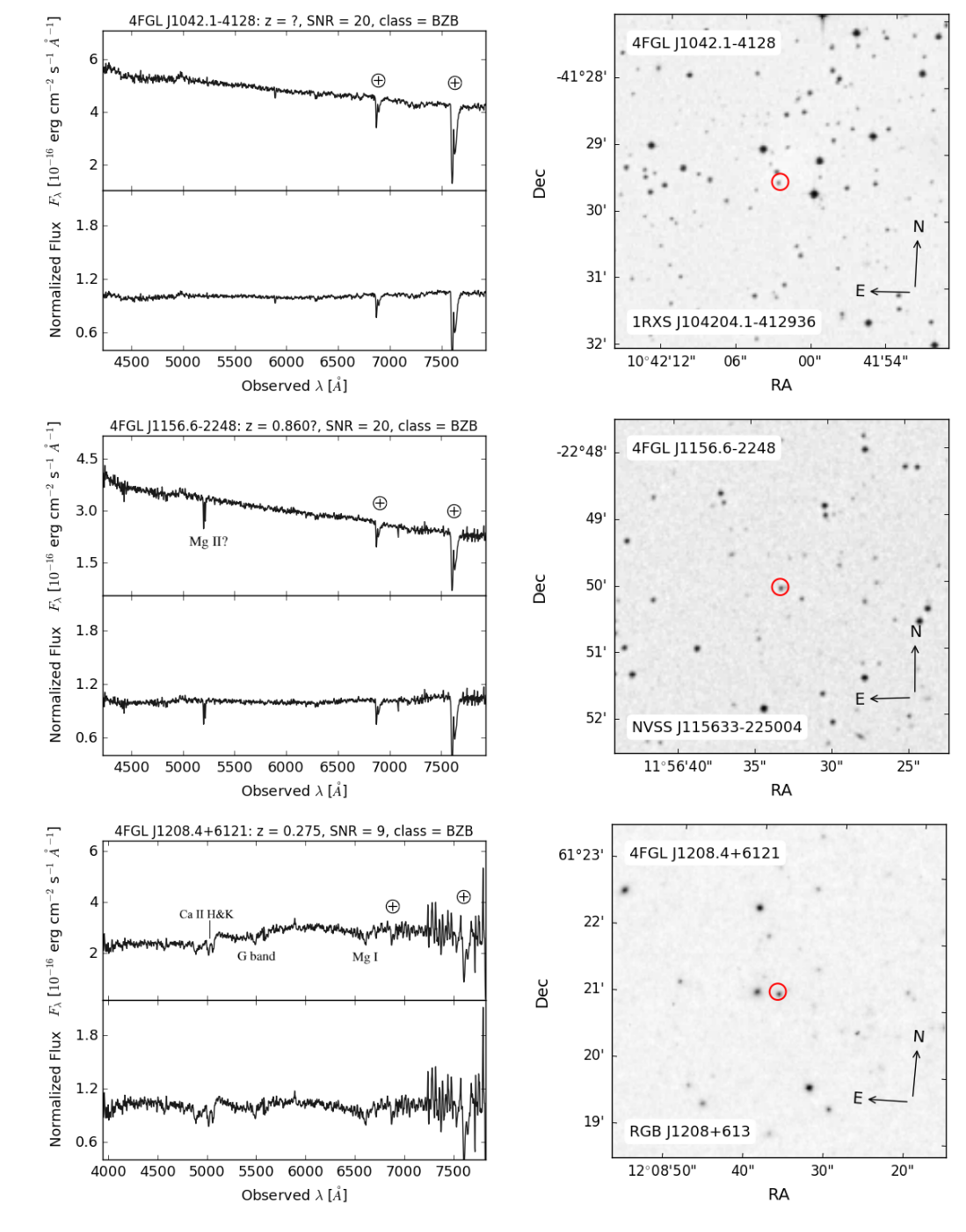}
    \caption{Continued from Figure 2.}
\end{figure*}

\begin{figure*}
%\ContinuedFloat
    \centering
    \includegraphics[scale=0.5]{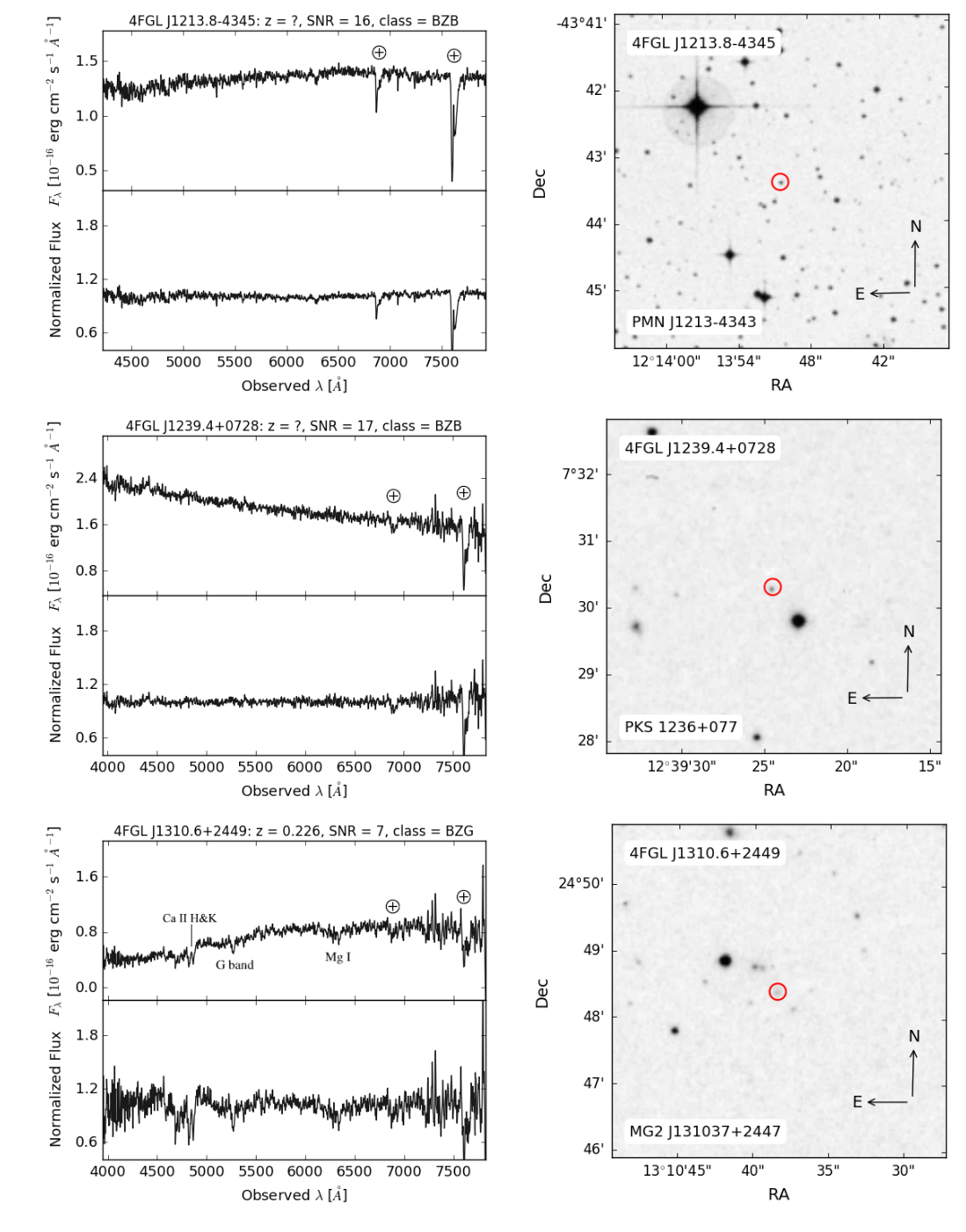}
    \caption{Continued from Figure 2.}
\end{figure*}

\begin{figure*}
%\ContinuedFloat
    \centering
    \includegraphics[scale=0.5]{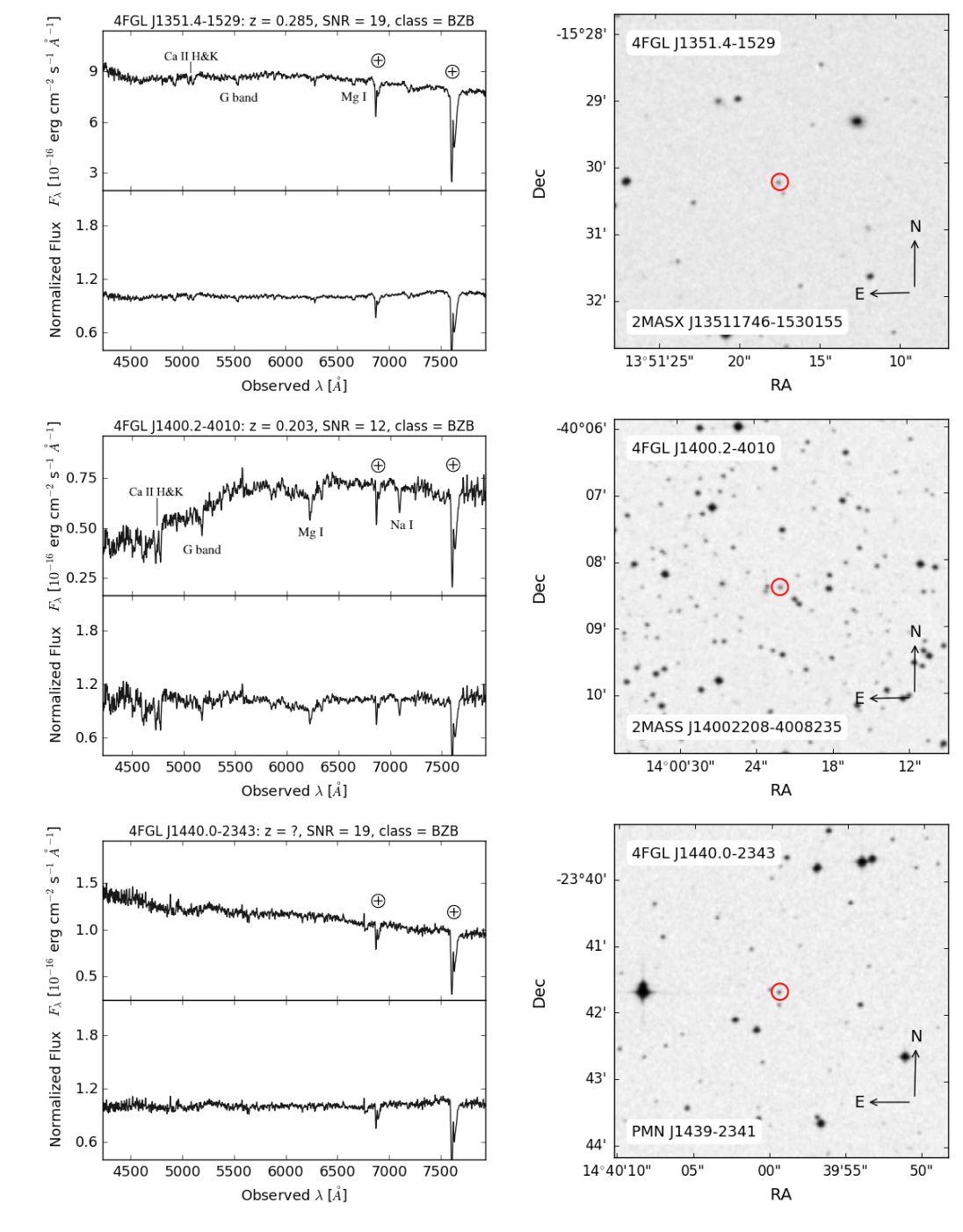}
    \caption{Continued from Figure 2.}
\end{figure*}

\begin{figure*}
%\ContinuedFloat
    \centering
    \includegraphics[scale=0.5]{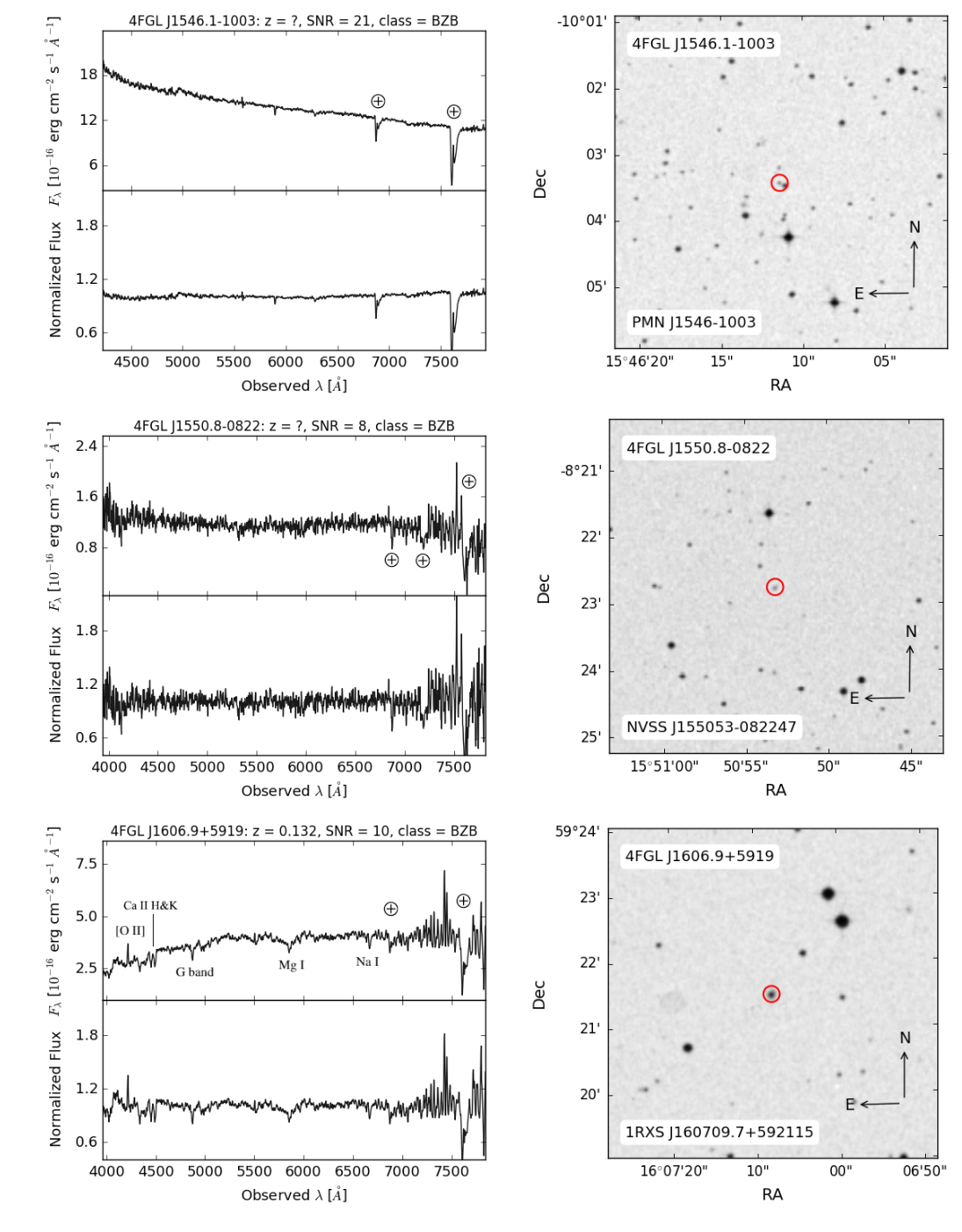}
    \caption{Continued from Figure 2.}
\end{figure*}

\begin{figure*}
%\ContinuedFloat
    \centering
    \includegraphics[scale=0.5]{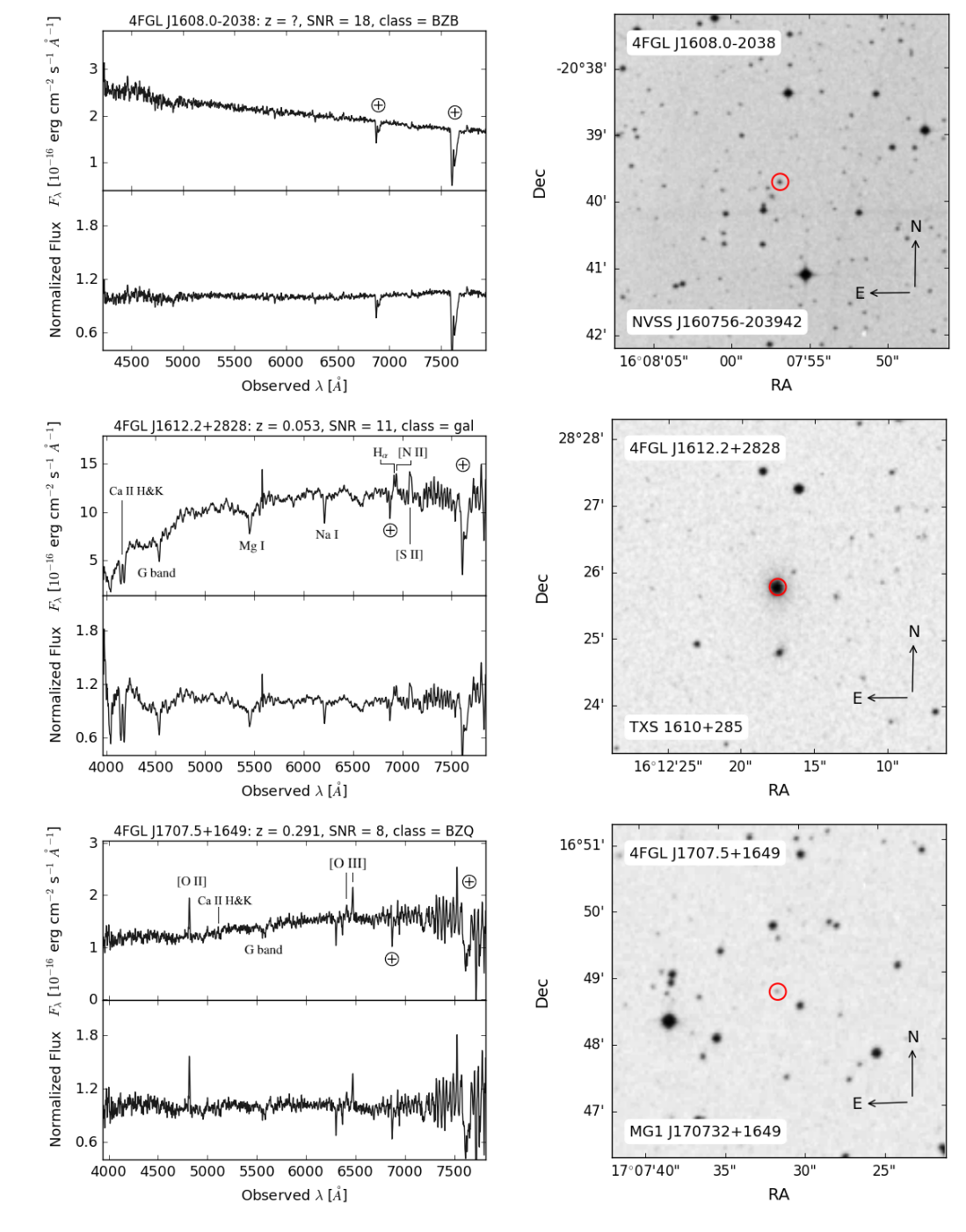}
    \caption{Continued from Figure 2.}
\end{figure*}

\begin{figure*}
%\ContinuedFloat
    \centering
    \includegraphics[scale=0.5]{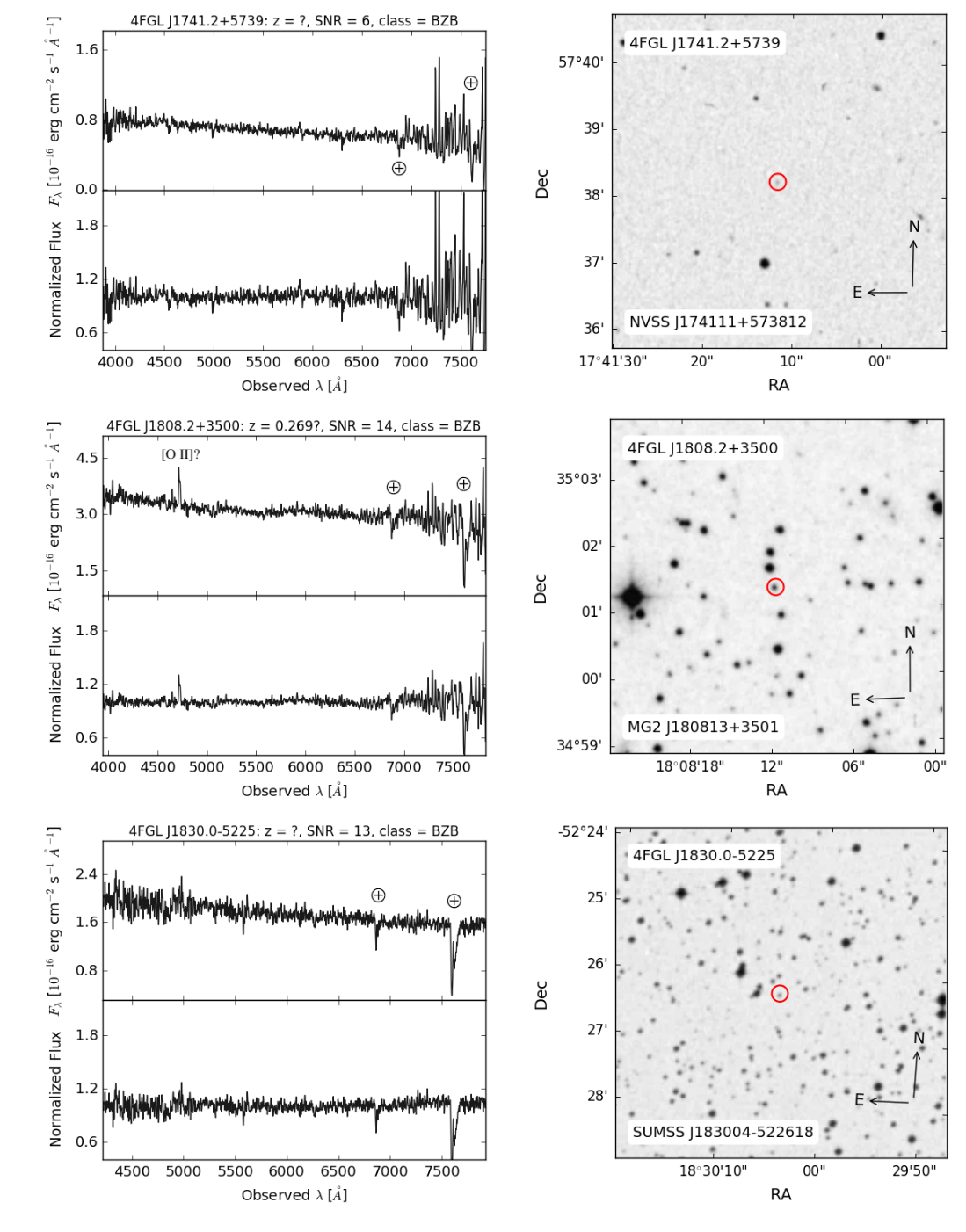}
    \caption{Continued from Figure 2.}
\end{figure*}

\begin{figure*}
%\ContinuedFloat
    \centering
    \includegraphics[scale=0.5]{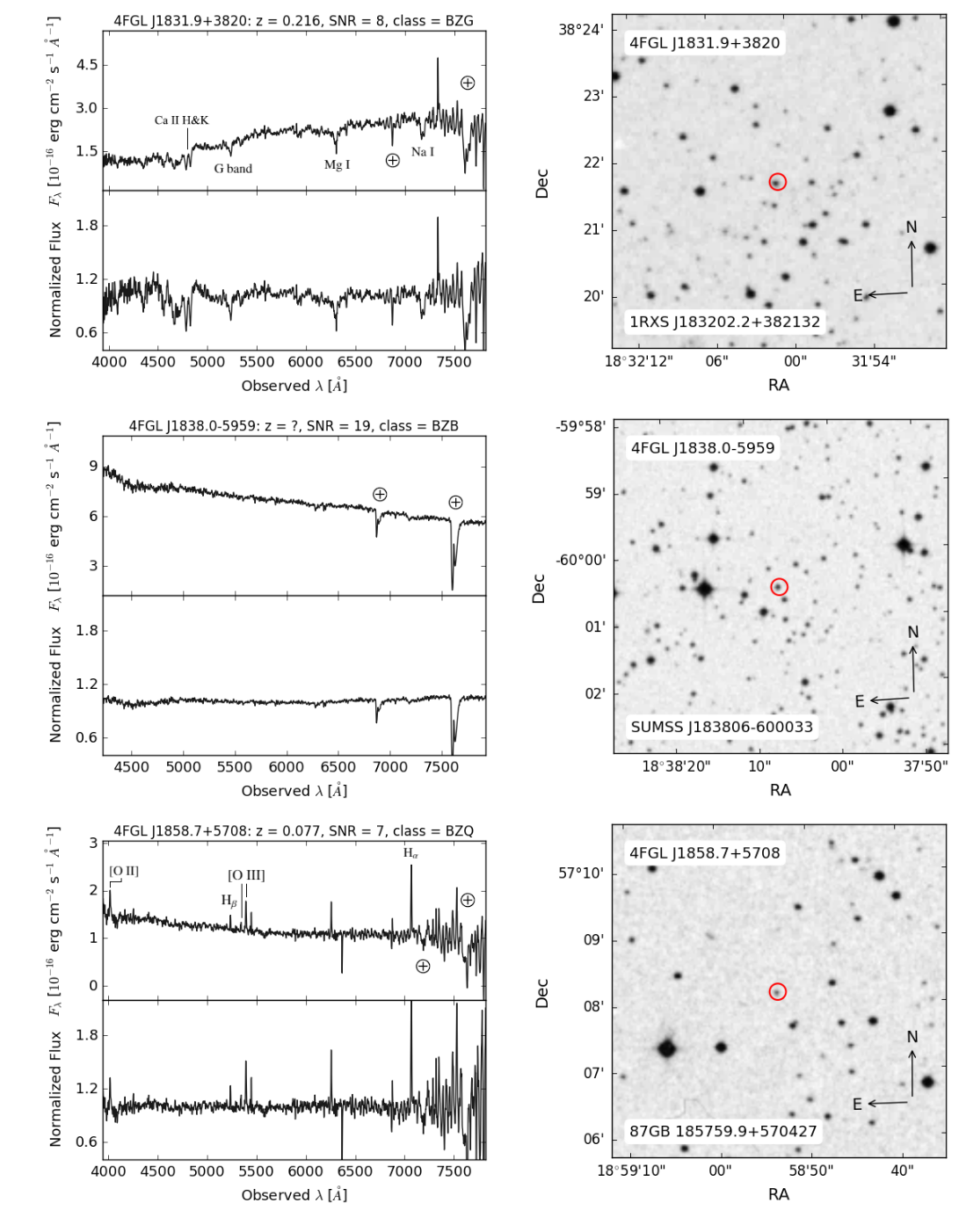}
    \caption{Continued from Figure 2.}
\end{figure*}

\begin{figure*}
%\ContinuedFloat
    \centering
    \includegraphics[scale=0.5]{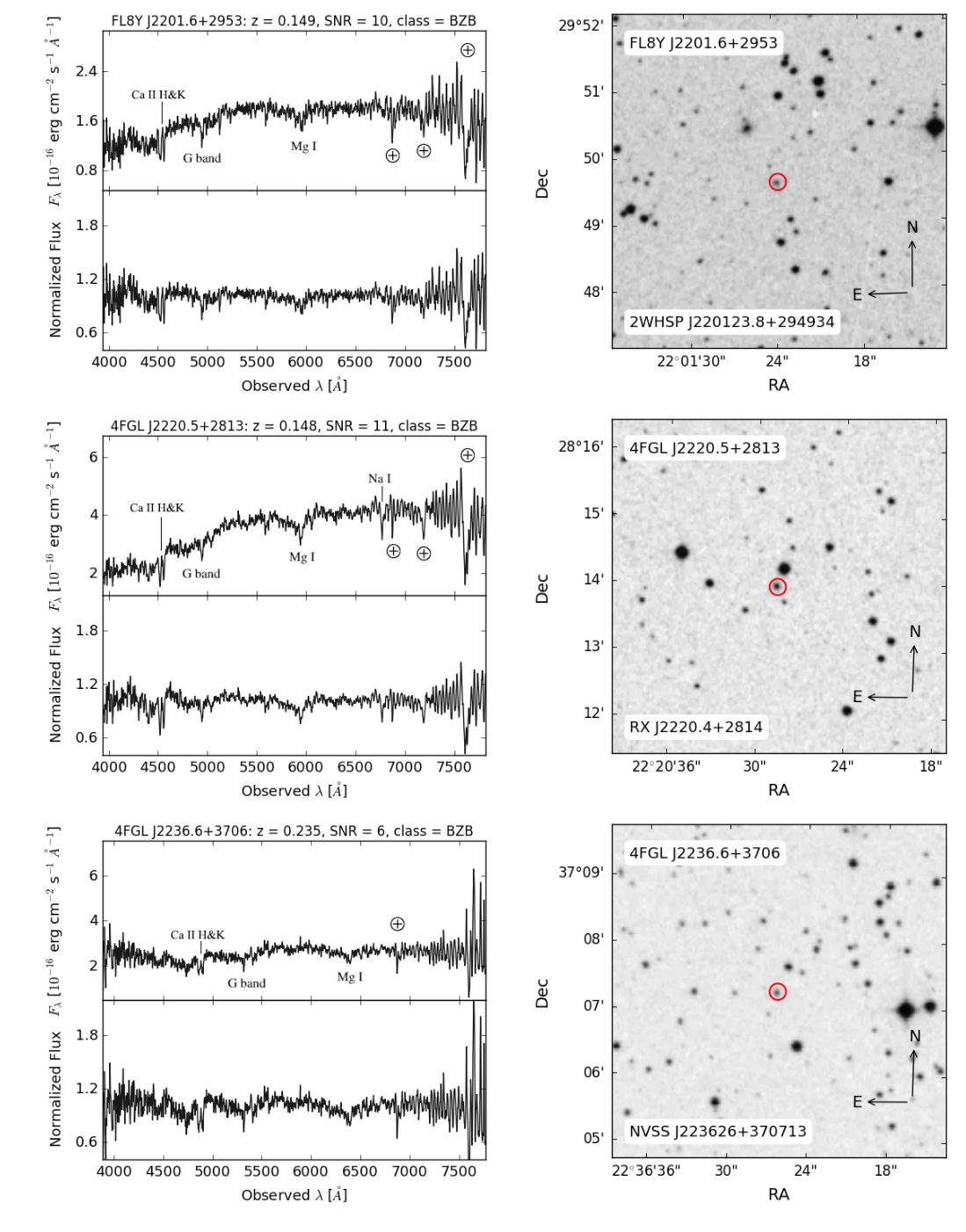}
    \caption{Continued from Figure 2.}
\end{figure*}

\begin{figure*}
%\ContinuedFloat
    \centering
    \includegraphics[scale=0.5]{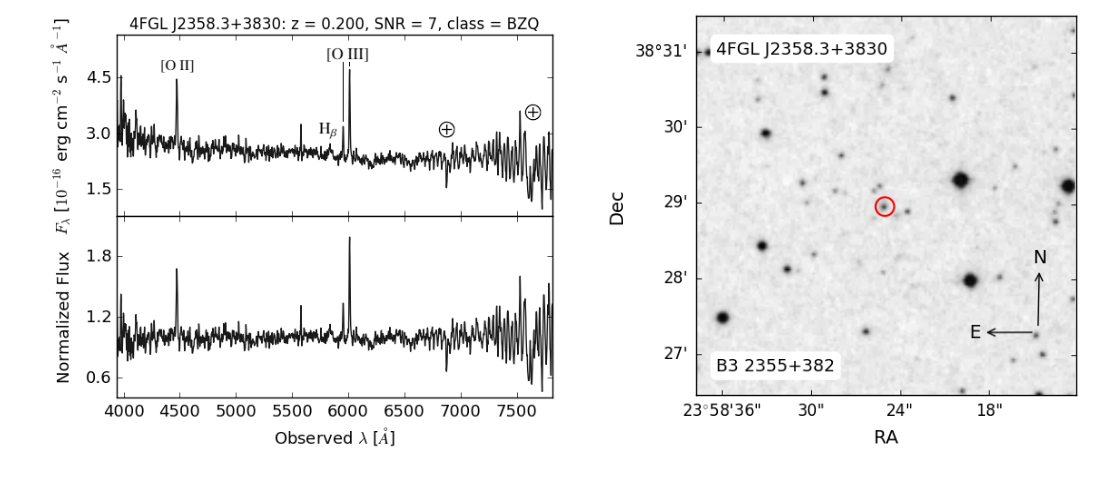}
    \caption{Continued from Figure 2.}
\end{figure*}

\end{document}